\documentstyle[aps,preprint,tighten,epsf,amssymb]{revtex}

\newcommand{\drv}[2]{\frac{d#1}{d#2}}

\def\stacking#1#2#3{\mathrel{\mathop{#3}\limits^{#1}\limits_{#2}}}

\begin{document}

\title{A Solution to the Gating Paradox.}
\author{L. P. Endresen and J. S. H{\o}ye\\}
\address{Institutt for fysikk, NTNU, N-7034 Trondheim, Norway}

\date{\today}
\maketitle

\vspace{-0.5cm}
\begin{abstract}
We introduce a Markov model for the gating of membrane channels. The
model features a possible solution to the so--called gating current
paradox, namely that the bell--shaped curve that describes the voltage
dependence of the kinetics is broader than expected from, and shifted
relative to, the sigmoidal curve that describes the voltage dependence
of the activation. The model also predicts some temperature dependence
of this shift, but presence of the latter has not been tested
experimentally so far.
\end{abstract}

\section*{INTRODUCTION}
\noindent
The gating of membrane channels is of vital importance for the
electrophysiological activity of nerve, heart and muscle. While some
of these channels appear to have fractal--like gating (Liebovitch,
1995), most membrane channels do display activity that can be well
approximated by a simple Markov process (Korn and Horn,
1988). However, Clay {\em et al.} (1995) revealed a gating current
paradox that has been difficult to explain with a standard type
(Hille, 1992) Markov model. The paradox is that the bell--shaped curve
that describes the voltage dependence of the kinetics is shifted
significantly relative to the sigmoidal curve that describes the
voltage dependence of the activation. The standard type models (Hille,
1992) does not allow such a shift. Also the former curve is broader
than the one predicted by the standard model.

Here we introduce a new Markov model, that extends and generalizes the
standard one. Our generalization consists of introducing an
alternative route between the open and the closed positions of the
gate. With two routes, or two membrane protein folding pathways, we
are able to obtain results consistent with the observed ones. Thus
such a model presents a possible resolution of the above paradox. A
more complete resolution requires investigation of the detailed
physical mechanism present in real membrane channels to see how they
compare with the model. The idea with two routes, a rapid one and a
slow one, is that the probability of choosing one or the other also
depends upon the voltage through a Boltzmann factor. This will affect
the kinetics, but not the equilibrium distribution (stationary state),
and a relative shift of curves can take place.

\vspace{-0.3cm}
\section*{THE MODEL}
\noindent
We imagine that a membrane channel has one open and one closed state
as in the simplest standard (Hille, 1992) Markov model for this
problem. However, between these states we now assume that there exist
two routes ($i = 1,2$). This gives,
\begin{equation}
\label{eq1}
\begin{array}{ccc}
  \hspace{0.25cm} & \stacking{{\alpha}_{1}}{{\beta}_{1}} {\rightleftharpoons} & \hspace{0.25cm} \vspace{-0.3cm}  \\
  \text{\huge C} \hspace{0.25cm} & & \hspace{0.25cm} \text{\huge O} \vspace{-0.5cm} \;, \\
  \hspace{0.25cm} & \stacking{{\alpha}_{2}}{{\beta}_{2}} {\rightleftharpoons} & \hspace{0.25cm}  
\end{array}
\end{equation} 

\noindent
where the rate constants ${\alpha}_1$, ${\alpha}_2$ and ${\beta}_1$,
${\beta}_2$, which are functions of voltage (but are constant at any
given voltage), control the transitions between the closed (C) and the
open (O) states of the gate. The ${\alpha}_i$ is the rate for a closed
channel to open, and ${\beta}_i$ the rate for an open channel to
close. We introduce effective rate constants $\alpha$ and $\beta$,
\begin{eqnarray}
\label{eq2}
{\alpha} &=& p_1 {\alpha}_1 + p_2 {\alpha}_2 \\
\label{eq3}
{\beta}  &=& p_1 {\beta}_1  + p_2 {\beta}_2 \;, 
\end{eqnarray}

\noindent
where the probabilities $p_1$ and $p_2$ are related in a standard way
to the difference $\Delta G_{b}$ in energy barriers that must be
overcome for each of the two routes,
\begin{eqnarray}
\label{eq4}
p_1 &=& \frac{\exp(-\frac{\Delta G_{b}}{2kT})}{\exp(\frac{\Delta G_{b}}{2kT}) + \exp(-\frac{\Delta G_{b}}{2kT})} \\
\label{eq5}
p_2 &=& \frac{\exp(\frac{\Delta G_{b}}{2kT})}{\exp(-\frac{\Delta G_{b}}{2kT}) + \exp(\frac{\Delta G_{b}}{2kT})} \;.
\end{eqnarray}

\noindent
Let $x$ denote the average fraction of gates that are open, or,
equivalently, the probability that a given gate will be open, and let
us imagine that a Markov (1906) model is suitable to describe the
gating. One then has, as usual
\begin{equation}
\label{eq6}
\drv{x}{t} = \alpha  (1-x) - \beta x
           = \frac{x_{\infty} -x}{\tau} \;,
\end{equation}
where
\begin{eqnarray}
\label{eq7}
	{x}_{\infty} &=& \frac{\alpha}{\alpha + \beta} \;, \\
\label{eq8}
	{\tau}   &=& \frac{1}{\alpha + {\beta} } \;.
\end{eqnarray}

\noindent
Here $x_{\infty}$ denotes the steady stationary state fraction of open
gates and ${\tau}$ the relaxation time. At equilibrium, the
probability for a channel to be in the open state is $x_{\infty}$, and
the probability to be in the closed state is $(1-x_{\infty})$.  The
ratio of these two probabilities is given by the Boltzmann
distribution,
\begin{equation}
\label{eq9}
\frac{x_{\infty}}{1-x_{\infty}} =
\exp\left(\frac{{\Delta G}_x}{kT}\right) \;,
\end{equation}

\noindent
where $T$ is the absolute temperature, $k$ is Boltzmann's constant,
and ${\Delta G}_x$ denote the energy difference between the open and
the closed positions. Thus,
\begin{equation}
\label{eq10}
{x}_{\infty} = \left(1+\exp\left[-\frac{{\Delta G}_x}{kT}\right] \right)^{-1} \;.
\end{equation}

\noindent
At equilibrium, each of the the forward reactions must occur just as frequently 
as each of the reverse reactions, giving,
\begin{equation}
\label{eq11}
\frac{{\alpha}_i}{{\beta}_i} = \exp\left(\frac{{\Delta G}_x}{kT}\right) \;.
\end{equation}

\noindent
This is the principle of detailed balance which is present in
dynamical systems (reversible mechanics). As in the standard model the
rates are then assumed to be,
\begin{eqnarray}
\label{eq12}
{\alpha}_i &=& \lambda_i  \exp\left(\frac{{\Delta G}_x}{2kT}\right) \\
\label{eq13}
{\beta}_i &=& \lambda_i  \exp\left(-\frac{{\Delta G}_x}{2kT}\right) \;,
\end{eqnarray}

\noindent
where $\lambda_i$ is assumed to be independent of ${\Delta G}_x$. Thus
the relaxation time (Eq. \ref{eq8}) can then be written as,
\begin{equation}
\label{eq14}
{\tau} = \frac{1}{{\alpha}+{\beta}} = \frac{1}{p_1 {\alpha}_1 + p_2 {\alpha}_2 + p_1 {\beta}_1  + p_2 {\beta}_2} \;.
\end{equation}

\noindent
Using Eqs. \ref{eq4}, \ref{eq5}, \ref{eq12} and \ref{eq13}, we obtain
\begin{eqnarray}
\label{eq15}
{\tau} &=& \frac{2\cosh(\frac{\Delta G_{b}}{2kT})}{({\alpha}_1 + {\beta}_1) \exp(-\frac{\Delta G_{b}}{2kT}) + ({\alpha}_2 + {\beta}_2) \exp( \frac{\Delta G_{b}}{2kT})} \\
\label{eq16}
       &=& \frac{\cosh(\frac{\Delta G_{b}}{2kT})}{ \cosh(\frac{{\Delta G}_x}{2kT}) \left[\lambda_1 \exp(-\frac{\Delta G_{b}}{2kT}) + \lambda_2 \exp(\frac{\Delta G_{b}}{2kT} \right]} \\
\label{eq17}
       &=& \frac{\cosh(\frac{\Delta G_{b}}{2kT})}{ \lambda \cosh(\frac{{\Delta G}_x}{2kT})
\left[\exp(-\frac{\Delta G_{b}}{2kT}-\gamma) + \exp(\frac{\Delta G_{b}}{2kT} + \gamma) \right]} \\
\label{eq18}
       &=& \frac{\cosh(\frac{\Delta G_{b}}{2kT})}{ 2 \lambda \cosh(\frac{{\Delta G}_x}{2kT}) \cosh(\frac{\Delta G_{b}}{2kT} + \gamma)} \;,
\end{eqnarray}

\noindent
where,
\begin{eqnarray}
\label{eq19}
{\gamma} &=& \frac{1}{2} \log \left(\frac{\lambda_2}{\lambda_1} \right) \\
\label{eq20}
{\lambda} &=& \sqrt{\lambda_1 \lambda_2} \;.
\end{eqnarray}

\noindent
To be more specific the voltage dependences of ${\Delta G}_x$ and
${\Delta G}_b$ are needed. For the energy difference between the open
state and the closed state we assume as usual,
\begin{equation}
\label{eq21}
{\Delta G}_x = G_{\rm closed} - G_{\rm open} \equiv q_x (v-v_x) - s_xT\;,
\end{equation}

\noindent
where the term $q_x v_x$ is due to the difference in mechanical
conformation energy between the two states; $q_x v$ represents the
electrical potential energy change associated with the redistribution
of charge during the transition, and $s_x$ is due to the difference in
entropy between the two states. A similar expression can be assumed
for the energy difference between the two barriers in route 1 and 2,
\begin{equation}
\label{eq22}
{\Delta G}_b = G_1 - G_2 \equiv q_b (v-v_b) - s_b T\;.
\end{equation}

\noindent
Here $v$ is voltage, while $q_x$, $v_x$, $s_x$, $q_b$, $v_b$, and
$s_b$ are constants. One notes that the curve for the relaxation time
$\tau$ has a shift in position due to the term $\gamma$. Inserted for
the special case ${\Delta G}_b = {\Delta G}_x$ the above yields,
\begin{eqnarray}
\label{eq23}
{x}_{\infty} &=& \left(1+\exp\left[\frac{-q_x (v-v_x) + s_xT}{kT}\right] \right)^{-1} \\
\label{eq24} 
{\tau} &=& \left(2 \lambda \cosh\left[\frac{-q_x(v-v_x) + T(s_x - 2\gamma k)}{2kT}\right] \right)^{-1} \;.
\end{eqnarray}

\noindent
Here we find that the voltage dependence of the curve for the
relaxation time (Eq. \ref{eq24}) is shifted by an amount $2\gamma k T
/ q_x$ relative to the steady state activation curve (Eq. \ref{eq23}),
which means that the magnitude of the shift depends upon
temperature. With ${\Delta G}_b \neq {\Delta G}_x$ expression
(Eq. \ref{eq24}) becomes more complex as follows from
(Eq. \ref{eq18}), and the shape of the former curve is modified. This
however, is dealt with in the next section.

\vspace{-0.3cm}
\section*{RESULTS}
\noindent
We now will compare the model with the experimental results of Clay
{\em et al.} (1995) and show that it is consistent with the latter.
Thus it presents a mechanism that represents a possible solution to
the gating current paradox. The temperature--dependence of the
currents were not considered in those experiments, so here $s_x$ and
$s_b$ can be incorporated into $v_x$ and $v_b$. With use of
Eqs. \ref{eq21} and \ref{eq22}, Eqs. \ref{eq10} and \ref{eq18} become,
\begin{eqnarray}
\label{eq25}
{x}_{\infty} &=& \frac{1}{1+\exp\left(\frac{v_x-v}{k_x}\right)} \\
\label{eq26} 
{\tau} &=& \frac{\cosh(\frac{v-v_b}{2k_b})}{ 2 \lambda \cosh(\frac{v-v_x}{2k_x}) \cosh(\frac{v-v_b}{2k_b} + \gamma)} \;,
\end{eqnarray}

\noindent
where $k_x = kT/q_x$ and $k_b = kT/q_b$. These expressions were
evaluated numerically adjusting the parameters present to obtain a
best possible fit to the experimental data. A least squares fit
weighting various points in accordance with experimental uncertainty
was used.  The results of this evaluation is shown in the figure below
where the data of Clay {\em et al.} (1995) is presented together with
the curves given by Eqs. \ref{eq25} and \ref{eq26} using the
parameters shown in the figure text.

\noindent
However, the curves are not very sensitive to the values of these
parameters except $\gamma$, so they can be varied quite a bit and
still give essentially the same curves. From these curves we find that
the model is fully consistent with the experimental results within the
uncertainties in the latter.

\vspace{-0.5cm}
\section*{DISCUSSION}
\noindent
We have presented a Markov model that yields a possible solution to
the gating current paradox announced by Clay {\em et al.}  (1995). It
gives a simple explanation of the voltage--shift of the bell--shaped
curve for the relaxation time relative to the steady state activation
curve. Also the width and shape of the relaxation time curve can be
modified in a way consistent with experiments. A novel feature of the
present model is that the voltage--shift is temperature dependent. It
is not clear whether such a temperature dependence can be observed
experimentally.

\begin{figure}[hbt]
\epsfxsize=1.0\textwidth \epsffile{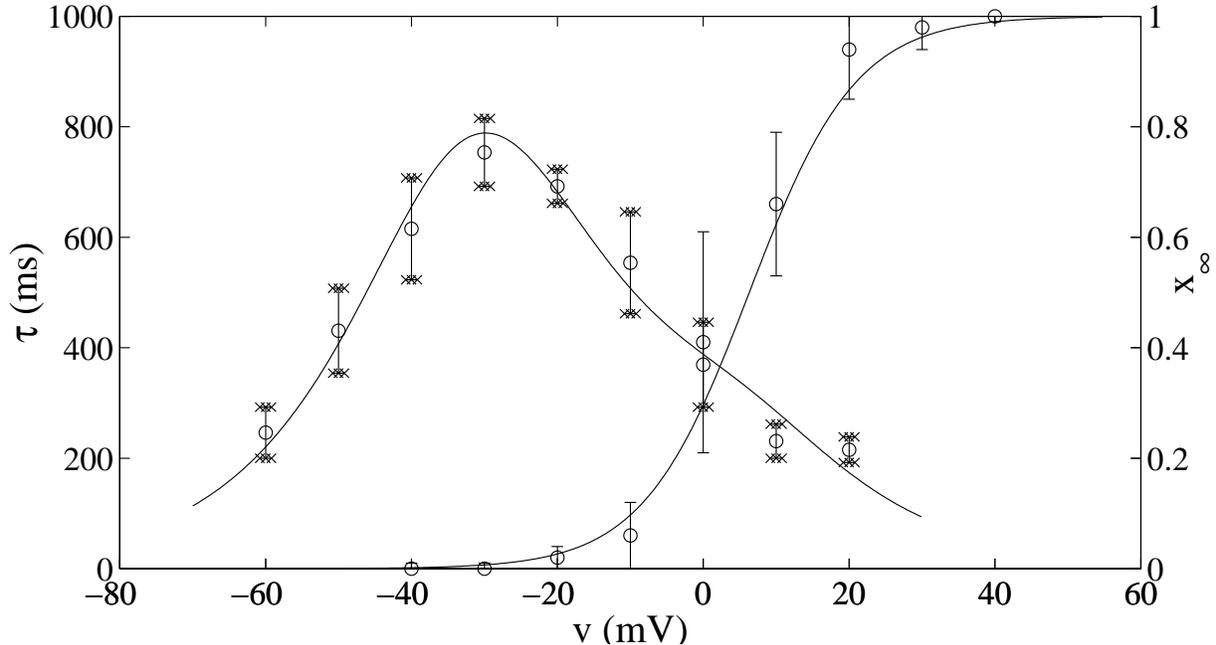}
\vspace{0.5cm}
\caption{\footnotesize The steady--state activation curve (Eq.
\ref{eq25}) and the bell--shaped curve for the relaxation time
(Eq. \ref{eq26}), with the parameters $v_x = 6.31 \, {\rm mV}$, $k_x =
7.31 \, {\rm mV}$, $\lambda = 0.31 \, {\rm s^{-1}}$, $v_b = -1.79 \,
{\rm mV}$, $k_b = 7.99 \, {\rm mV}$ and $\gamma = 1.89$. The error
bars indicate the $\pm$ mean standard deviation from the six
experiments of Clay {\em et al.}  (1995).}
\end{figure}
\vspace{0.3cm}

\noindent
{\footnotesize Lars Petter Endresen would like to thank professor Jan
Myrheim for illuminating discussions in connection with this
work. This work was supported with a fellowship from NTNU.}

\vspace{-0.3cm}
\section*{REFERENCES}
\begin{description}
\item {Clay, J. R., A. Ogbaghebriel, T. Paquette,
B. I. Sasyniuk, and A. Shrier.} 1995. A quantitative description of
the E--4031--sensitive repolarization current in rabbit ventricular
myocytes. {\em Biophysical Journal.} 69:1830--1837.
\item{Hille, B.}  1992. Ionic channels of excitable
membranes. Sunderland, Massachusetts. 485--490.
\item {Korn, S. J., and Horn, R.} 1988. Statistical discrimination 
of fractal and Markov models of single--channel gating. 
{\em Biophysical Journal.} 54:871--877.
\item{Liebovitch, L. S.}  1995. Single channels: from
Markovian to fractal models. {\em In} Cardiac Electrophysiology: from
Cell to Bedside. D. P. Zipes, and J. Jalife, editors. Philadelphia:
Saunders. 293--304.
\item{Markov, A. A.}  1906. Extension de la loi de
grands nombres aux {\'e}v{\'e}nements dependants les uns de autres.
{\em Bulletin de La Soci{\'e}t{\'e} Physico--Math{\'e}matique de
Kasan.}  15: 135--156.
\end{description}

\end{document}